\documentclass[aps,prb,twocolumn,groupedaddress,showpacs]{revtex4-1}
\usepackage{amssymb,amsmath,graphicx}

\newcommand{\dimalpha}{\tilde{\alpha}}
\newcommand{\dimbeta}{\tilde{\beta}}

\begin{document}

\title{Polarized light emission from individual incandescent carbon nanotubes}


\author{S. B. Singer}

\author{Matthew Mecklenburg}

\author{E. R. White}

\author{B. C. Regan}
\affiliation{%
Department of Physics and Astronomy, University of California, Los Angeles, California, 90095\\
and California NanoSystems Institute, Los Angeles, California, 90095
}


\date{June 10, 2011}

\begin{abstract}
We fabricate nanoscale lamps which have a filament consisting of a single multiwalled carbon nanotube.  After determining the nanotube geometry with a transmission electron microscope, we use Joule heating to bring the filament to incandescence, with peak temperatures in excess of 2000~K.  We image the thermal light in both polarizations simultaneously as a function of wavelength and input electrical power. The observed degree of polarization is typically of the order of 75\%, a magnitude predicted by a Mie model of the filament that assigns graphene's optical conductance $\pi e^2/2 h$ to each nanotube wall.
\end{abstract}


\maketitle

Understanding the optical properties of nanoscopic objects remains an outstanding problem in physics.  An object's geometric and electronic structures, which at the nanoscale are interrelated, determine these properties completely.  The polarization of light absorbed, emitted, or scattered by an object reports on both, and can reveal information not otherwise available optically, \cite{1983Bohren} \emph{e.g.}, the axis of symmetry of an unresolved emitter.  Thus polarization provides a handle on understanding the structures key for manipulating and controlling electromagnetic fields at the nanoscale.\cite{2009Bharadwaj}

Carbon nanotubes are theoretically tractable, nontrivial emitters available in a wide array of morphologies, making them ideal test objects for a study comparing structure with polarization.  From a size standpoint the two limiting cases have been extensively investigated.  Previously polarization-dependent absorption, emission, or reflection has been reported from macroscopic carbon nanotube aggregates such as films,\cite{2004Islam,2005Murakami} fibers, \cite{2000Hwang} bundles, \cite{2003LiPol} and arrays. \cite{2004Wang, 2009Cubukcu} Individual single-walled carbon nanotubes (SWCNTs) have also shown polarization effects in their emission, \cite{2003Hartschuh,2004Lefebvre,2007Mann,2010Moritsubo}  scattering, \cite{2002Jorio,2004Sfeir,2010Joh} and absorption. \cite{2004Lefebvre,2007BerciaudNano}  

As a class of material, individual multiwalled carbon nanotubes (MWCNTs) straddle the boundary between the macroscopic and molecular limits.  A many-layered MWCNT has a dielectric response like that of bulk graphite, while decreasing the number of layers to the limiting value of one produces a SWCNT  -   essentially a large molecule with discrete spectral features.  Polarization effects in the challenging intermediate regime represented by individual MWCNTs have not been explored previously.

We present here a study of polarized light emission from individual incandescent MWCNTs, where the emitter peak temperatures are in excess of 2000~K.   Our data quantify the polarization and its wavelength dependence throughout the visible into the near infrared (450--1100 nm), with sub-wavelength resolution of the emitter in both polarizations simultaneously. Also unique, however, is that each individual emitter is imaged with atomic resolution in a transmission electron microscope (TEM).  These images fully characterize the source,  providing precise determinations of the nanotube lengths, radii, and numbers of walls. The MWCNTs observed have 10-20 walls, core radii of $\sim 2$~nm, and outer radii of 7--9 nm, putting them in the regime where molecular spectral features are not expected and the classical theory provides a reasonable first approximation. Thus we analyze the observed  degree of optical polarization ($DoP$) using the Mie model of a small, conducting tube and the structural information provided by the TEM.

\begin{figure}\begin{center}
\includegraphics[width=.95 \columnwidth]{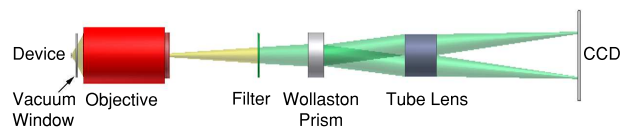}
\caption{\label{fig:optics}(Color online) Microscope with a 10~nm bandpass color filter and Wollaston prism in the infinity space.  The 100$\times$ microscope objective (numerical aperture NA$=0.5$) corrects for the aberration introduced by the vacuum window and achieves a Gaussian width resolution $r=0.21 \lambda/\text{NA} = 0.42 \lambda$. }
\end{center}\end{figure}

\begin{figure}\begin{center}
\includegraphics[width=.95 \columnwidth]{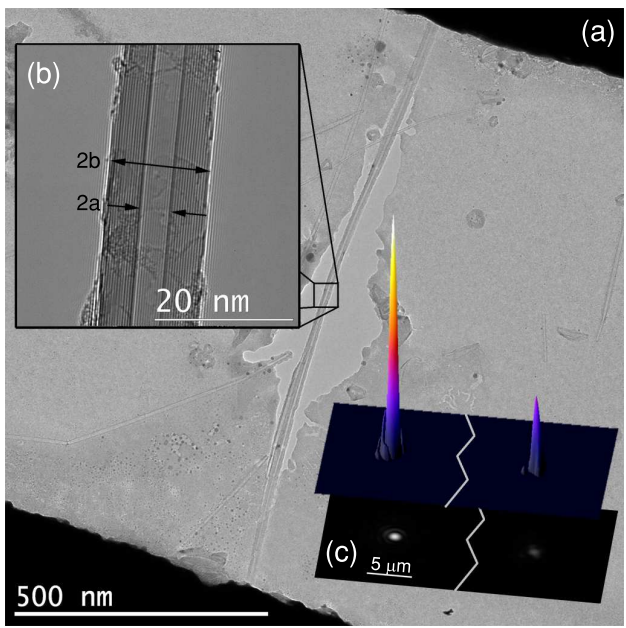}
\caption{\label{fig:neferTEM} (Color online) TEM and optical images of a nanotube device.  The TEM image (a) shows a device after optical data acquisition, where the nanotube is fully suspended for much of the distance between the two metal contacts. The membrane surrounding the nanotube has disintegrated, indicating that it reached temperatures above 2000~K (Ref.~\onlinecite{2007Begtrup}). A high magnification TEM image (b) shows the inner ($a=2.5$~nm) and outer ($b=7.4$~nm) radii of the nanotube with its $14\pm 1$ walls. Two incandescent spots, one for each polarization, are seen in a representative $\lambda=900$~nm optical image (c) from the same device.  Most of the dark space separating the two polarizations has been cut for display purposes.}
\end{center}\end{figure}

The nanotube lamp fabrication process and the basics of the optical system have been described previously.\cite{2009Fan}  The two-terminal devices have an active element consisting of an arc-discharge grown MWCNT suspended on an electron-transparent Si$_3$N$_4$ membrane window in a 2~mm$\times$2~mm$\times 0.2$ mm silicon chip.  After imaging a device in the TEM, we load it into a high-vacuum chamber for optical characterization (see Fig.~\ref{fig:optics}) and apply an increasing bias voltage over a period of some minutes.  With sufficient applied power the membrane neighboring the midpoint of the device disintegrates, as indicated by a decrease in the current at constant voltage accompanied by a 10\%--20\% increase in device brightness.  Both effects are attributed to the improved thermal isolation, and therefore higher temperature, of the center section of the nanotube.  Taking the membrane disintegration to initiate near the dissociation temperature of Si$_3$N$_4$, we adopt 2000~K as a conservative lower bound on the central temperature of the device at that applied bias voltage.\cite{2007Begtrup} After the optical experiments have been completed the membrane disintegration is confirmed in the TEM (see Fig.~\ref{fig:neferTEM}).  The Si$_3$N$_4$ membrane is a thin, excellent insulator with a correspondingly tiny ($\lesssim 10^{-15}$) emissivity in the visible; we have not seen evidence of light emission from any source other than the nanotube itself.

Simultaneous imaging of the nanotube in both polarizations is achieved by placing a Wollaston prism in the optical microscope's infinity space (see Fig.~\ref{fig:optics}). The quartz prism introduces an angular displacement of  $\sim 2^\circ$ between the two orthogonal polarizations, which creates  an effective position displacement of 7~mm (70~$\mu$m) in the image (object) plane. Both single polarization images of the radiating nanotube are easily captured by the $13.3\times13.3$~mm$^2$ CCD sensor.  With the prism mounted on a rotation stage and an array of 10~nm bandpass filters we image both polarizations simultaneously as a function of power, wavelength, and prism angle.
 
We define parallel ($||$) and perpendicular ($\perp$) light polarizations by the prism orientation that maximizes the contrast in a single image, and calculate the $DoP$ of the nanotube's emission with \cite{1983Bohren}
\begin{equation}
DoP = \frac{\dot{N}^{||} - \dot{N}^{\perp}}{\dot{N}^{||} + \dot{N}^{\perp}}.
\end{equation}
To most closely approximate an isothermal field of view, $\dot{N}$ is taken to be the photon count rate in the spot's brightest pixel, an effective area of ($130$~nm)$^2$. $\dot{N}$ depends on $\lambda$, the emitting area $A$, the temperature $T$ via the Planck factor, and the optical system efficiency, \cite{2009Fan} but these common factors cancel in the $DoP$, leaving 
\begin{equation}
DoP(\lambda,T) = \frac{\int_{\text{NA}} Q_{abs}^{||}(\lambda,T) d\Omega - \int_{\text{NA}} Q_{abs}^\perp(\lambda,T) d\Omega}{\int_{\text{NA}} Q_{abs}^{||}(\lambda,T)  d\Omega + \int_{\text{NA}} Q_{abs}^\perp(\lambda,T) d\Omega}.
\end{equation}
Here $Q_{abs}$ is the absorption efficiency, or equivalently emissivity, \cite{1983Bohren,2009Schuller} of the nanotube in the specified polarization.

\begin{figure}\begin{center}
\includegraphics[width=.95 \columnwidth]{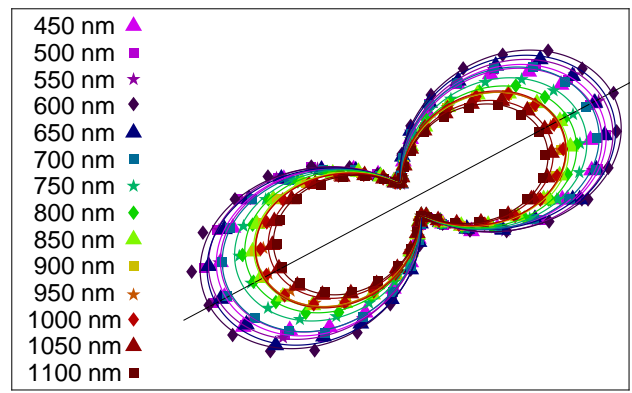}
\caption{\label{fig:clover} (Color online) Polar plots of spot intensity as a function of prism angle $\theta$ for  $\lambda= 450$--$1100$~ nm in 50 nm increments. The radial coordinate is proportional to the photon count rate in the spot's brightest pixel, normalized by its minimum intensity for display purposes.   The straight line indicates the orientation of the nanotube. Connecting the data for each wavelength are fits to the function $A\cos^2 (\theta+\theta_0) + B$.}
\end{center}\end{figure}

Figure \ref{fig:clover} shows representative data from the nanotube device in Fig.~\ref{fig:neferTEM}, plotted as a polar function of intensity and prism angle.    The phase offset $\theta_0$ gives the angle of maximum polarization, which matches the orientation of the nanotube's long axis as determined by the relative orientations of the optical and TEM images. \cite{2009Fan} A 2D Gaussian fit to the nanotube's optical image alone also yields a major axis consistent with the other two determinations.  The $DoP$ for this device is seen to vary with wavelength, showing a gentle peak near $\lambda =600$~nm.

Polarization effects in the absorption or emission of light from carbon nanotubes, especially SWCNTs, have been treated theoretically. \cite{2003Milosevic,2004Saito,2005Goupalov}  Emission perpendicular to the nanotube axis is thought to be strongly suppressed by the ``depolarization effect," \cite{1994Ajiki,1995Benedict,2003Marinopoulos} where charge induced on the nanotube walls cancels the internal electric field. \cite{1994Ajiki,2004Saito}  However, we are unaware of theory predicting the $DoP$ expected from a single MWCNT. 

For a model with which we can compare our experimental results, we turn to classical Mie theory, which has found success in predicting the scattering and absorption of gold nanoparticles\cite{2006vanDijk} and metallic dust grains\cite{2008Rosenberg}.  We model the nanotube as an infinitely long, cylindrical tube with inner radius $a$ and outer radius $b$.  The problem of scattering and absorption by a solid infinite right circular cylinder has been treated by Bohren and Huffman \cite{1983Bohren}.  Their derivation considers a plane wave of wavevector $k=2 \pi/\lambda$ incident on the cylinder at an angle $\zeta$ measured from the tube axis, and solves for the extinction and scattering by calculating the scattered wave and invoking conservation of energy.  The scattered fields are described with an infinite series of coefficients $a_n$ and $b_n$ that weight the solutions to the wave equation in cylindrical coordinates.  Requiring continuity of the electromagnetic fields on the surface fixes the series coefficients.  Modifying this derivation for the tubular case, we find extinction and scattering efficiencies 
\begin{equation}
\begin{split}
Q_{ext,\perp} &= \frac{4}{k(a+b)}\,\mbox{Re}\left[a_0^{\perp} + 2 \sum_{n = 1}^\infty a_n^{\perp}\right] \quad \text{and}\\
Q_{sca,\perp} &= \frac{4}{k(a+b)}\left[|a_0^{\perp}|^2 + 2 \sum_{n = 1}^\infty (|a_n^{\perp}|^2+|b_n^{\perp}|^2)\right],
\end{split}
\end{equation}
for an incident wave polarized perpendicular to the tube axis.  Similar relations hold for the parallel extinction and scattering efficiencies $Q_{ext,||}$ and $Q_{sca,||}$, which are found by making the substitutions $a_n^{\perp} \rightarrow b_n^{||}$ and $b_n^{\perp} \rightarrow a_n^{||}$ in the relations for $Q_{ext,\perp}$ and $Q_{sca,\perp}$. For a narrow ($a,b \ll \lambda$) tube in vacuum,  the $a_n$ and $b_n$ coefficients are, to the lowest nontrivial order in $\dimalpha \equiv ka \ll 1$ and $\dimbeta \equiv kb \ll 1$,
\begin{equation}
\begin{split}
b_0^{||} &= \frac{-i \pi}{4} (\dimbeta^2 - \dimalpha^2)(m^2 - 1) \sin^2 \zeta, \\
b_1^{||} &= \frac{-i \pi}{4} \frac{\dimbeta^2 (\dimbeta^2 - \dimalpha^2) (m^4 - 1)}{\dimbeta^2 (m^2 + 1)^2 - \dimalpha^2 (m^2 - 1)^2} \cos^2 \zeta, \,\text{and}\\
a_1^\perp &= \frac{-i \pi}{4} \frac{\dimbeta^2 (\dimbeta^2 - \dimalpha^2) (m^4 - 1)}{\dimbeta^2 (m^2 + 1)^2 - \dimalpha^2 (m^2 - 1)^2}.
\end{split}
\end{equation}
The tube's complex index of refraction $m$ is assumed to satisfy $|m|\dimalpha,|m|\dimbeta \ll 1$ here.  The absorption efficiency $Q_{abs} = Q_{ext} - Q_{sca}\simeq Q_{ext}$, since $Q_{sca}$ is negligible for $\dimalpha, \dimbeta \ll 1$.  In terms of the dielectric constant $\epsilon = m^2$,
\begin{equation}\label{eq:Qans}
\begin{split}
Q_{abs}^{||} &= \pi (\dimbeta - \dimalpha) \big(  \text{Im}[\epsilon - 1]\sin^2 \zeta  \\
&  + 2 \dimbeta^2 \text{Im}\left[ \frac{\epsilon^2 -1}{\dimbeta^2 ( \epsilon + 1)^2 - \dimalpha^2 (\epsilon - 1)^2}\right]\cos^2 \zeta \big) , \\
Q_{abs}^{\perp} &= 2 \pi (\dimbeta - \dimalpha) \dimbeta^2 \text{Im}\left[ \frac{\epsilon^2 -1}{\dimbeta^2 ( \epsilon + 1)^2 - \dimalpha^2 (\epsilon - 1)^2}\right]. 
\end{split}
\end{equation}
As expected, $Q_{abs}^{\perp}$ is independent of incidence angle, and at $\zeta = 0$ (incident wave $k$ parallel to the tube axis), the absorption efficiencies are equal.

The dielectric constant can be written in the form $\epsilon = 1 + i \frac{4 \pi \sigma_{\text{3D}}}{\omega} = 1 + i \frac{\sigma_{\text{3D}} Z_0}{k}$, where the conductivity $\sigma_{\text{3D}}$ is in general complex and $Z_0=4 \pi/c \simeq 377~ \Omega$ is the impedance of free space.  To complete the model, we approximate each nanotube wall\cite{2005Murakami,2010Skulason} as having the frequency-independent, two-dimensional (2D) optical conductance per square $\sigma_g=\pi e^2/2 h=\pi \alpha/Z_0$ expected for a graphene sheet\cite{2002Ando,2006GusyninPRL,2010Tree} ($\alpha=e^2/\hbar c\simeq 1/137$ is the fine-structure constant).  We define an effective three-dimensional (3D) conductivity $\sigma_{\text{3D}} \equiv \sigma_g/\delta = n \pi \alpha/ Z_0 (b-a)$, where  $b - a$ is the  thickness, $n$ is the number of walls, and $\delta\simeq 0.34$~nm is their spacing.  The absorption coefficients are then determined in terms of the dimensionless quantity $s \equiv \sigma_{\text{3D}} Z_0 (b-a)= n \pi \alpha$. 

Our calculations are based on the expressions (\ref{eq:Qans}), but to elucidate their structure we consider the zero thickness limit where $\dimalpha,\dimbeta \rightarrow x = kr$.  In this limit, the prior assumption $|m|x \ll 1$ is equivalent to $s x \ll 1$ and  Eq.~\ref{eq:Qans} reduces to
\begin{equation}\label{eq:zerothick}
\begin{split}
Q_{abs}^{||} &= \pi s \left[\sin^2 \zeta + \frac{2 x^2 \cos^2 \zeta}{s^2 + 4 x^2} \right], \quad \text{and}\\
Q_{abs}^{\perp} &= \pi s \frac{2 x^2}{s^2 + 4 x^2}.
\end{split}
\end{equation}

Equation~(\ref{eq:zerothick}) gives a $DoP$ between  $1/3$ for $s/x$ small and unity for $s/x$ large at normal incidence $(\zeta=\pi/2$). This trend with tube radius $r$ is consistent with data showing that SWCNTs have greater $DoP$s than our MWCNTs, with sub-nanometer diameter SWCNTs exhibiting the greatest $DoP$s. \cite{2003Hartschuh,2004Lefebvre,2007Mann} A quantitative treatment of SWCNTs is beyond the scope of this theory, since we do not account for resonance or selection rule effects. \cite{1994Ajiki,1995Benedict,2003Marinopoulos,2003Milosevic,2004Saito,2005Goupalov} Generally Eq.~(\ref{eq:Qans}) gives smaller $DoP$s for large tubes with large cores ($b-a \ll b$), and $DoP$s  largely independent of $b$ for small cores ($a\ll b$).

\begin{figure}\begin{center}
\includegraphics[width=.95 \columnwidth]{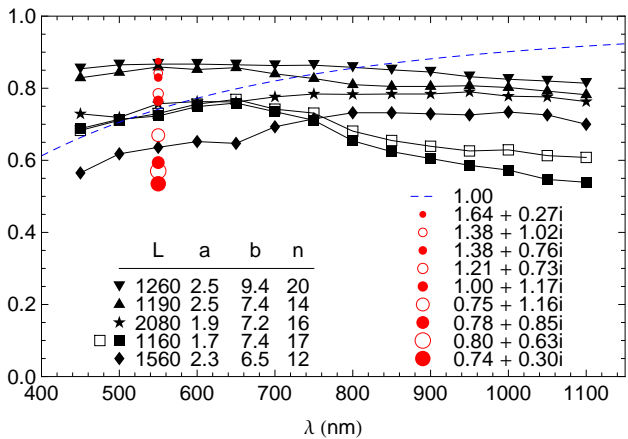}
\caption{\label{fig:polarization}(Color online) Degree of polarization as a function of wavelength.  The sequences of black symbols represent the measured $DoP$ for five nanotube devices.  Their geometric parameters as determined by TEM are given in the inset table on the left, where $L$, $a$, and $b$ are given in nanometers and the number of walls $n$ has an error of $\pm 1$.  Data from the device of Figs.~\ref{fig:neferTEM} and \ref{fig:clover} is represented with $\blacktriangle$.  The curves $\blacksquare$ and $\square$ were acquired from one device at input powers of 451 and 362~$\mu$W respectively.  The dashed curve is calculated from Eq.~(\ref{eq:ourexperiment}) using the radii $a$ and $b$ of the $\blacktriangle$ device, and assuming graphene's wavelength-independent conductance $\sigma_g=\pi \alpha/Z_0$ per wall --- there are no adjustable parameters.  The red symbols are the expected $DoP$ assuming the same geometry for other, generally complex, measured values of the per layer conductance (in units of $\sigma_g$) of multilayer graphite as summarized in Ref.~\onlinecite{2010Skulason}.  These values are only valid at $\lambda=550$~nm.}
\end{center}\end{figure}
Returning to (\ref{eq:Qans}), we integrate the absorption efficiencies over the NA to find the observed $DoP$,
\begin{equation}\label{eq:ourexperiment}
\begin{split}
&DoP = \\
&\frac{ s^4 (\dimbeta+\dimalpha)^2 + 4 s^2 \dimbeta^2 (\dimbeta^2+\dimalpha^2)}{s^4 (\dimbeta+\dimalpha)^2  + 4 s^2 \dimbeta^2 (\dimbeta^2+\dimalpha^2) (2 G + 1) + 32 \dimbeta^4 (\dimbeta-\dimalpha)^2 G}
\end{split}
\end{equation}
where $G = \int_{\text{NA}} d\Omega/\int_{\text{NA}} \sin^2 \zeta d\Omega= 1.07$ for our NA$=0.5$.

Figure \ref{fig:polarization} summarizes the $DoP$ data for five different nanotube devices.  Also shown is the prediction of Eq.~(\ref{eq:ourexperiment}) for nanotube radii as determined by TEM and  a graphene-like optical conductance of $\sigma_g$ per wall.  The prediction varies little for the range of $a$'s and $b$'s explored, but is sensitive to the conductance; taking other values from the literature can move the $DoP$ sufficiently to encompass all of the data. Thus we find that Mie theory and optical constants characteristic of multilayer graphene reproduce the observed magnitude of the $DoP$.

Beyond the evident coarse agreement with the Mie model, the data raise some questions. First, it is not yet clear how to explain the observed variance between devices with the detailed geometric information made available by the TEM images.  Second, for a frequency-independent optical conductivity, Mie theory predicts a $DoP$ that increases with wavelength;  much of the data indicate the opposite trend.  Furthermore, the $DoP$ is generally observed to decrease with increasing applied bias (or equivalently temperature), with the effect stronger at longer wavelengths.  Figure~\ref{fig:polarization} shows a second curve (labeled $\square$), representing a midpoint temperature $\sim 400$~K cooler, for the  device that illustrates this effect most dramatically.  At maximum bias these devices are operating at temperatures where optical and zone-boundary phonon modes ($\hbar \Omega \simeq 0.16~e$V$ \simeq 1900$~K) are thermally populated; we suggest that the resulting decrease in the electronic mean free path makes the bias electric field less effective in establishing the polarization axis \cite{2000Yao}.  However, a more comprehensive theoretical picture than our Mie model will be required to capture these effects.


In conclusion, we have structurally characterized individual MWCNTs, brought them to incandescence, and measured their polarized light emission.  Assuming optical conductances characteristic of graphene, a simple model based on classical Mie theory predicts the magnitude of the observed degree of polarization of $\sim 75\%$.  A complete explanation of the variations around this value that depend on  wavelength, device configuration, and device temperature requires a more sophisticated model.  Future research on MWCNTs, especially those with fewer walls, will further probe the classical-quantum transition regime and advance our understanding of nanoscale optical devices.

This project is supported by NSF CAREER Grant No. 0748880.  All TEM work was performed at the Electron Imaging Center for NanoMachines at UCLA.


\bibliography{polarization}

\end{document}